\documentclass{ws-procs975x65}

\begin{document}

\title{REPORT ON SESSION QG3 OF THE 15TH MARCEL GROSSMANN MEETING}

\author{JORGE PULLIN}

\address{Department of Physics and Astronomy, Louisiana State
  University, Baton Rouge, LA 70803\\
pullin@lsu.edu}

\author{PARAMPREET SINGH}

\address{Department of Physics and Astronomy, Louisiana State
  University, Baton Rouge, LA 70803\\
psingh@lsu.edu}

\begin{abstract}
We summarize the talks presented at the QG3 session (loop quantum
gravity: cosmology and black holes) of the 15th Marcel Grossmann
Meeting held in Rome, Italy on July 1-7 2018.
\end{abstract}



\bodymatter

\vskip1cm

The session was devoted to results on black holes and cosmology in the
context of loop quantum gravity. It was a lively session where
several new developments were presented in 10 talks by a mixture of beginning
and established researchers. Two of the talks, by Guillermo
Mena-Marug\'an and Anzhong Wang were 30 minutes talks. The rest of the talks were 15 minutes each. Most of the talks focused on investigating phenomenologically interesting implications of loop quantum gravity in the early universe and black hole scenarios. And, some developments such as pertaining to the non Gaussianities in loop quantum cosmology could lead to potentially observational signatures. In the following, we provide a brief summary of various results in the order they were presented during the session.

Guillermo Mena--Marug\'an kicked off the session talking about perturbations in loop quantum cosmology,
a joint work with Beatriz Elizaga de Navascu\'es and Daniel Mart\'{\i}n de Blas. The focus of the talk was a comparison between the ``hybrid'' and the ``dressed metric'' approaches to understand inhomogeneities in loop quantum cosmology. In both the approaches, perturbative modes can be described by a collection of harmonic oscillators with time-dependent mass. However, masses differs if one uses the ``hybrid'' or  the ``dressed metric'' approach. The difference in mass is found to originate from the difference in canonical versus effective approach to compute  derivatives in the tensor mass. At the bounce, the mass is positive in the hybrid approach in a sector that contains all kinetically dominated solutions, but is negative for the dressed metric for those solutions. It was argued that this negativity may be an obstruction to the definition of adiabatic states. The mass properties are found to be important for dynamics and constrain the permissible choices of vacuum states.  More details can be found in \cite{ElizagaNavascues:2017avq}.

Beatriz Elizaga de Navascu\'es talked about joint work with Guillermo Mena--Marug\'an and S. Prado Loy on fermionic perturbations treated in the hybrid approach. There the homogeneous isotropic background cosmology is quantized using loop quantum cosmology techniques and the fermionic inhomogeneous perturbations using traditional Fock quantization. The back reaction of the perturbations was studied. For further details, see 
\cite{ElizagaNavascues:2018zzu}.

Suzana Bedi\'c discussed joint work with Gregory Vereshchagin about the probability of inflation in loop quantum cosmology. They develop an effective phase space and a Liouville measure on it after reviewing several measures that have been proposed in literature. Contrary to the case of traditional general relativity where different choices of initial surface change significantly the likelihood of inflation, in loop quantum cosmology there are as set of preferred solutions that constitute an attractor in the contracting phase. Speaker discussed possible contradiction with previous results in loop quantum cosmology. Details can be found in   \cite{Bedic:2018gqu}.

Killian Martineau described work with Aur\'elien Barrau, J. Grain and S. Schander on trans Planckian effects in loop quantum cosmologies. In particular the duration of inflation was shown to be well constrained if initial conditions are set in the classical contracting phase and the inflaton potential is confining. In a second part, the dressed metric approach is compared to the deformed algebra approach. The trans Planckian problem arises if one has more than 60 e-folds since the observed modes become much smaller than the Planck scale at the bounce. More information in   \cite{Martineau:2017sti,Martineau:2017tdx}.

Sreenath Vijayakumar presented joint work with Ivan Agull\'o and Boris Bolliet discussing non-Gaussianities  in dressed metric approach to perturbations in loop quantum cosmology. It was found that non-Gaussianity generated in loop quantum cosmology has a shape that allows  large  non-Gaussianity at  low  multipoles, being at the same time compatible with observational  constraints. Further, observations allow to put a constraint on the value of the inflaton at the bounce.  More details in 
\cite{Agullo:2017eyh}.

Anzong Wang talked on joint work with Bao-Fei li and Parampreet Singh on the qualitative dynamics of pre-inflationary universe in loop quantum gravity. Three different variants of loop quantum cosmology were considered and although the post bounce dynamics is similar in all three the pre bounce differs in one of them. In particular, existence of quantum deSitter branch was demonstrated in one modification of loop quantum cosmology which can potentially result from loop quantum gravity. The existence of inflationary attractors is shown for all the cases for a variety of potentials. All three variants reproduce general relativity for late times. Details are discussed in    \cite{Li:2018opr,Li:2018fco}.

Flavio Bombacigno discussed work in collaboration with F. Cianfrani and G. Montani on making the Immirzi parameter a dynamical field and its application in a minisuperspace model leading to a bounce scenario. They also investigate the use of the Immirzi parameter as a relational time. More information in   \cite{Bombacigno:2016siz}.

Gioele Botta spoke about joint work with Emanuele Alesci and Gabriele V. Stagno in which they introduce a new regularization scheme for quantum cosmology within loop quantum gravity based on the tools of quantum reduced loop gravity. It leads to a new effective dynamics. It has been explored for Bianchi I space-times and there are plans to extend it to Bianchi IX model. More details in   \cite{Alesci:2017kzc}.

Leonardo Modesto described joint work with Qiqi Zhang and Cosimo Bambi on singularity free black holes in conformal gravity. Two black hole solutions are obtained via a conformal rescaling of the Kerr metric, one with a regular and one with a singular space-time. Geodesic evolution is studied. More can be found in   \cite{Zhang:2018qdk}.

Flora Moulin discussed joint work with Aur\'elien Barrau, Killian Martineau and Julien Grain. The talk consisted in two parts. The first was about a bouncing black hole in which a black hole tunnels into a white hole and examined the resulting fluxes as possible explanation of fast radio bursts. More details of this part in
\cite{Barrau:2018kyv}. The second part was loop black holes cross section scalar and fermionic
fields were studied on the background of the non-singular black hole of Modesto. The black hole cross section diminishes due to the quantum effects. More details in \cite{Moulin:2018uap}.

\section*{Acknowledgments}

This work was supported in part by grant NSF-PHY-1454832 and NSF-PHY-1603630, funds of the Hearne Institute for Theoretical
Physics and CCT-LSU. 




\end{document}